\documentclass[multi]{cambridge6A}
\usepackage[numbers]{natbib}
\usepackage{rotating}
\usepackage{floatpag}
\rotfloatpagestyle{empty}
\usepackage{amsthm}
\usepackage{graphicx}
\usepackage{amsmath}
\usepackage{amsfonts}
\usepackage{ubec}
\usepackage{amssymb}

\begin{document}

\newcommand{\bs}{\boldsymbol}

\author{\vskip1cm}
\affiliation{}
\author{Nathan Goldman}
\affiliation{Center for Nonlinear Phenomena and Complex Systems, Universit\'e Libre de Bruxelles, CP 231, Campus Plaine, B-1050 Brussels, Belgium}
\author{Nigel Cooper}
\affiliation{T.C.M. Group, Cavendish Laboratory, J.J. Thomson Avenue, Cambridge CB3 0HE, United Kingdom}
\author{Jean Dalibard}
\affiliation{Laboratoire Kastler Brossel, Coll\`ege de France, CNRS, ENS-PSL Research University, UPMC-Sorbonne Universit\'es, 11 place Marcelin Berthelot, 75005, Paris, France}

\chapter[Preparing and probing Chern bands with cold atoms]{Preparing and probing Chern bands \\with cold atoms}

\null
\vskip4cm
\begin{abstract}
The present Chapter discusses methods by which topological Bloch bands  can be prepared in cold-atom setups. Focusing on the case of Chern bands for two-dimensional systems, we describe how topological properties can be triggered by driving atomic gases, either by dressing internal levels with light or through time-periodic modulations. We illustrate these methods with concrete examples, and we discuss recent experiments  where geometrical and topological band properties have been identified.
\end{abstract}

\section{Introduction}

Ultracold atoms constitute a promising physical platform for the preparation and  exploration of novel states of matter \cite{Bloch2008,Dalibard2011,GoldmanReview}. In particular, the engineering of topological band structures with cold-atom systems, together with the capability of tuning interactions between the particles, opens an interesting route towards the realization of intriguing strongly-correlated states with topological features, such as fractional topological insulators and quantum Hall liquids \cite{Cooper2008}. 

This Chapter is dedicated to the preparation and the detection of topological band structures characterized by non-zero Chern numbers \cite{Bernevig}. Such  \emph{Chern bands}, which constitute the building blocks for realizing (fractional) Chern insulators \cite{Parameswaran2013816,bergholtzliu}, arise in 2D systems presenting time-reversal-symmetry (TRS) breaking effects. For instance, non-trivial Chern bands naturally appear in the Harper-Hofstadter model \cite{Hofstadter1976}, a lattice penetrated by a uniform flux, where they generalize the  (non-dispersive) Landau levels to the lattice framework. Additionally, Chern bands also appear in staggered flux configurations, such as in Haldane's honeycomb-lattice model  \cite{Haldane1988}, or in lattice systems combining Rashba spin-orbit coupling and Zeeman (exchange) fields. 

The atoms being charge neutral, ``magnetic" fluxes cannot be simply produced by subjecting optical lattices to ``real" magnetic fields. It is the aim of this Chapter to review several schemes that have been recently implemented in laboratories with the goal of realizing synthetic magnetic fields leading to Chern bands for cold atoms. The Chapter is structured as follows: Section \ref{Dalibard_Chern} describes how the Chern number is related to physical observables defined in a lattice framework. In particular, it clarifies the link between recent Chern-number measurements performed in cold bosonic gases and the more conventional (electronic) quantum Hall effect. Section \ref{Dalibard_models} reviews a few lattice models, and it relates their properties to the well-known Landau levels of the continuum. The main Section \ref{Dalibard_Implementation} describes diverse schemes by which Chern bands can be prepared and probed for cold atoms; a special emphasis is placed upon experimentally realized schemes. The last Section \ref{Dalibard_Discussion} is devoted to final remarks and discussions.

\section{Bloch waves, the anomalous velocity \\ and the Chern number}\label{Dalibard_Chern}

The Chern number -- a topological invariant $\nu_{\mathrm{ch}} \in \mathbb{Z}$ classifying fibre bundles -- naturally enters the description of particles moving in two-dimensional lattices, where it offers an elegant interpretation for the (anomalous) quantum Hall effect \cite{Thouless1982,Kohmoto:1985,Haldane1988}. This Section relates this topological invariant to physical observables and discusses methods to measure it in experiments.

\subsection{Bloch bands, the anomalous velocity and the Chern number}

The general problem of a particle subjected to a 2D space-periodic potential $U(\bs{r}+\bs{a})\!=\!U(\bs{r})$ starts by invoking Bloch's theorem, which stipulates that the eigenstates of the system can be decomposed as $\psi_{\lambda \bs{k}} (\bs{r}) \!=\! \exp (i \bs{r} \cdot \bs{k}) \, u_{\lambda \bs{k}} (\bs{r}) $, where $u_{\lambda \bs{k}} (\bs{r})$ has the periodicity of the lattice $u_{\lambda \bs{k}} (\bs{r}+\bs{a})=u_{\lambda \bs{k}} (\bs{r})$ and where $\bs{k}=(k_x,k_y)$ is the quasi-momentum. The associated eigenenergies $E_{\lambda} (\bs{k})$ display bands, labeled by the index $\lambda$, over the first Brillouin zone (FBZ) of the quasi-momentum $\bs{k}$. In the absence of additional potentials, a state $u_{\lambda \bs{k}} $ in a given Bloch band $\lambda$ with quasi-momentum $\bs{k}$ is characterized by the averaged velocity $ \bs{v}_{\lambda} (\bs{k})\!\equiv \! \langle u_{\lambda \bs{k}} \vert \hat{\bs{v}} \vert u_{\lambda \bs{k}} \rangle \!=\! (1/\hbar) \partial E_{\lambda} (\bs{k})/ \partial \bs{k}$, where $\hat{\bs{v}}$ denotes the velocity operator. 

\begin{figure}[h!]
\begin{center}
\includegraphics[width=12.5cm]{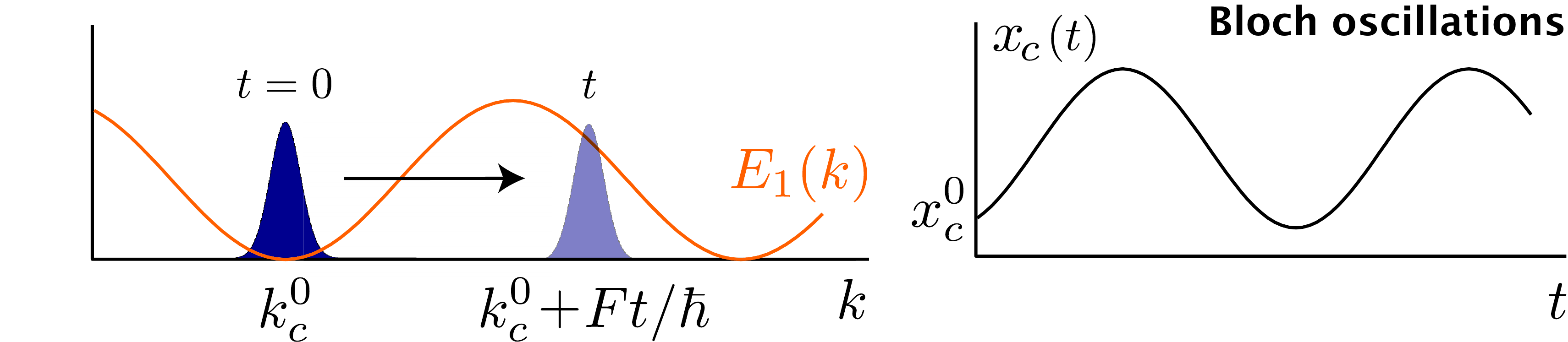} 
\caption{Bloch oscillations for a wave packet initially prepared in the lowest band $E_{1} $, with center-of-mass position ${x}_c^0$ and momentum $k_c^0$. The trajectory $x_c(t)$ results from the semi-classical equations of motion (\ref{Goldman_eq1}). }
\label{Goldman_fig1}
\end{center}
\end{figure}

Subjecting the lattice system to a constant force $F$ generates Bloch oscillations. Indeed, restricting the dynamics to 1D for now, and considering for simplicity a  semi-classical approach, the system is  described by the equations of motion
\begin{equation}
\hbar \dot {x}_c (t)= \hbar v_{\lambda} (k_c) = \frac{\partial E_{1} ({k}_c)}{\partial {k}_c} , \quad \hbar \dot {k}_c (t)= F,\label{Goldman_eq1}
\end{equation}
where ${x}_c(t)$ and $\hbar {k}_c(t)$ denote the center-of-mass position and momentum of a wave packet prepared in the lowest band $E_{1} $ (i.e. $\lambda=1$), see Fig.~\ref{Goldman_fig1}. The equations of motion (\ref{Goldman_eq1}) are valid when the force $F$ is weak enough to preclude any inter-band transitions.

When considering 2D lattices, an interesting effect adds to the standard Bloch oscillations, which involves the Berry curvature of the band \cite{Xiao2010}. Indeed, applying a force along a given direction, e.g. $\bs{F}=F_y \bs{1}_y$, modifies the averaged velocity $\bs{v}_{\lambda} (\bs{k})$ along the transverse ($x$) direction according to
\begin{eqnarray}
v_{\lambda}^x (\bs{k}) &=&  \frac{\partial E_{\lambda} (\bs{k})}{\hbar \partial k_x} - \frac{F_y}{\hbar} \Omega_{\lambda} (\bs{k}) , \\ 
\quad \Omega_{\lambda} (\bs{k}) &=& i \left ( \left< \partial_{k_{x}} u_{\lambda \bs{k}} | \partial_{k_{y}} u_{\lambda \bs{k}} \right> - \left< \partial_{k_{y}} u_{\lambda \bs{k}} | \partial_{k_{x}} u_{\lambda\bs{k}} \right> \right ),\label{Goldman_eq2}
\end{eqnarray}
where $\Omega_{\lambda} (\bs{k})$ denotes the Berry curvature of the band $\lambda$. Thus, the velocity in a state $u_{\lambda} (\bs{k})$  has two contributions: the usual \emph{band} velocity, responsible for Bloch oscillations [Fig.~\ref{Goldman_fig1}], and the so-called \emph{anomalous} velocity, which can produce a net drift transverse to the applied force. As for Eq.~(\ref{Goldman_eq1}), the general result (\ref{Goldman_eq2})  assumes the absence of inter-band transitions (i.e.~weak-force regime). 

The anomalous (transverse) velocity in Eq.~(\ref{Goldman_eq2}) can be isolated and observed experimentally by canceling any contribution from the band velocity, which can otherwise dominate.  This can be achieved by comparing trajectories for opposite forces, $\pm F_y$, noting that the anomalous velocity changes sign under reversal of the force \cite{Price:2012}. Alternatively, the average anomalous velocity can be isolated by uniformly populating the bands, namely by averaging the velocity in Eq.~(\ref{Goldman_eq2}) over the entire FBZ, since
\begin{equation}
\int_{\mathrm{FBZ}} \bigl ( \partial E_{\lambda} (\bs{k})/\partial k_x \bigr )  \, d^2 k =0 ,\label{Goldman_eq3}
\end{equation}
due to the periodicity of the energies in $k$-space. 

We now compute the total transverse velocity in the case of uniformly populated bands using Eq.~(\ref{Goldman_eq2}); explicit physical implementations will be discussed in the next Sections \ref{fermions}-\ref{thermal}. In the following, we consider a general square lattice system of size $A_{\mathrm{syst}} \!=\!L_x \!\times\! L_y$, characterized by a unit cell of size $A_{\mathrm{cell}} = d_x a \times d_y a$, where $a$ denotes the primitive lattice spacing\textsuperscript{1}\footnotetext[1]{For instance,  $d_x\!=\!d_y/2\!=\!1$ for the brick-wall lattice \cite{Tarruell2011}, and $d_x\!=\!d_y\!=\!2$ for the magnetic unit cell of a square lattice penetrated by a uniform flux $\Phi\!=\!\pi/2$ per plaquette \cite{Aidelsburger:2014}.}. The number of states within each band is $N_{\mathrm{states}}\!=\!A_{\mathrm{syst}}/A_{\mathrm{cell}}$. We write the total number of particles as $N_{\mathrm{tot}}=\sum_{\lambda} N^{(\lambda)}$, where $N^{(\lambda)}$ is the number of particles occupying a given band $\lambda$. We now make the assumption that each band is populated homogeneously, so that the average number of particles in a state $u_{\lambda \bs{k}} $ is uniform over the FBZ, and it is given by
\begin{equation}
\rho^{(\lambda)} (\bs{k})=\rho^{(\lambda)}=N^{(\lambda)} / N_{\mathrm{states}}.\label{Goldman_eq4}
\end{equation}
Using Eqs. (\ref{Goldman_eq2}) and (\ref{Goldman_eq4}), we obtain the total averaged velocity along the  direction transverse to the force
\begin{eqnarray}
&&v^x_{\mathrm{tot}}=  \sum_{\lambda} \rho^{(\lambda)} \sum_{\bs{k}} v_{\lambda}^x (\bs{k})  =  - \frac{F_y A_{\mathrm{cell}}}{h} \sum_{\lambda} N^{(\lambda)}  \, \nu_{\mathrm{ch}}^{(\lambda)}  , \label{Goldman_eq5} \\
&& \nu_{\mathrm{ch}}^{(\lambda)}= \frac{1}{2 \pi} \sum_{\bs{k}} \Omega_{\lambda} (\bs{k}) \, \Delta k_{x}\Delta k_{y} \longrightarrow \frac{1}{2 \pi} \int_{\mathrm{FBZ}} \Omega_{\lambda} (\bs{k}) \, d^2 k \in \mathbb{Z}, \label{Goldman_eq6}
\end{eqnarray}
where $\Delta k_{x,y}=2 \pi/L_{x,y}$. The latter equations reveal that the total velocity is related to the quantities $\nu_{\mathrm{ch}}^{(\lambda)}$, which converge towards the Chern number of the bands $\lambda$ when taking the thermodynamic limit $L_{x,y}\rightarrow \infty$. The Chern number $\nu_{\mathrm{ch}}^{(\lambda)}$ is an integer, obtained by averaging the Berry curvature $\Omega_{\lambda} (\bs{k})$ over the FBZ [Eq.~(\ref{Goldman_eq6})]; it is a topological invariant, meaning that $\nu_{\mathrm{ch}}^{(\lambda)}$ remains a constant as long as the spectral gaps to other bands do not vanish, see Refs. \cite{Thouless1982,Kohmoto:1985}. As announced above, any contribution from the band velocity cancels under the homogeneous-population condition (\ref{Goldman_eq4}), as a direct consequence of Eq. (\ref{Goldman_eq3}). \\

\begin{figure}[h!]
\includegraphics[width=11.5cm]{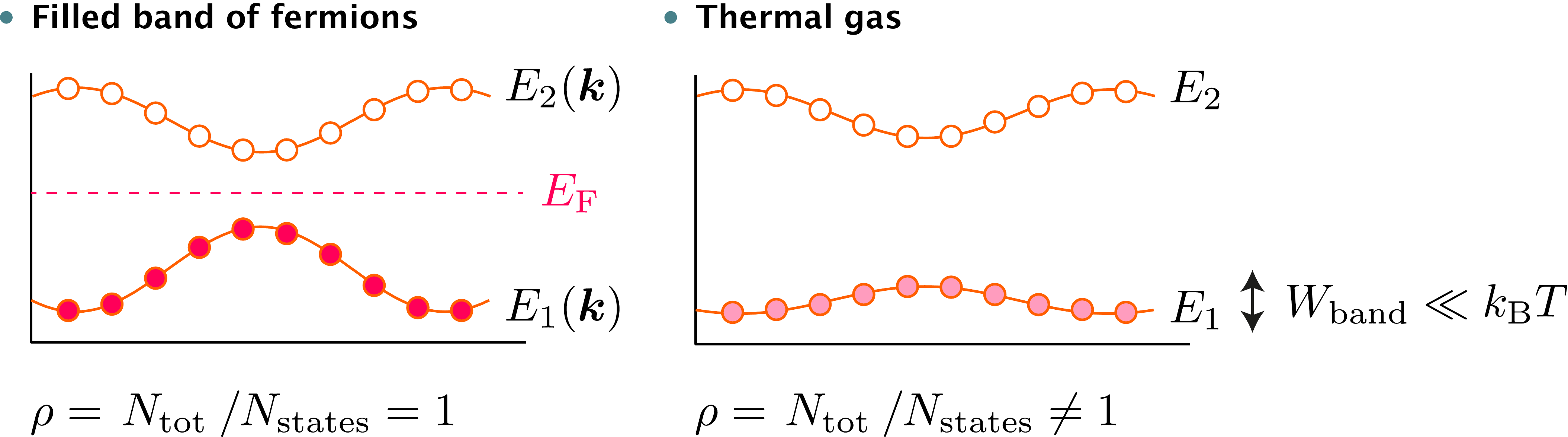} 
\caption{Two realizations of uniformly populated energy bands, suitable to reveal the Chern number in experiments. (left) Considering fermions at zero temperature, the band is perfectly filled by setting the Fermi energy within the spectral gap. (right) A system of bosons uniformly populate the band when the temperature is large compared to the bandwidth  $W_{\mathrm{band}}$, but small compared to the gap.}
\label{Goldman_fig2}
\end{figure}

The following Sections \ref{fermions}-\ref{thermal} discuss two different physical realizations that revealed the Chern numbers $\nu_{\mathrm{ch}}^{(\lambda)}$ in experiments, through the homogeneous population of energy bands [Fig.~\ref{Goldman_fig2}]. These Sections aim to clarify the link between the quantum Hall effect \cite{Thouless1982,Kohmoto:1985,Bernevig}, as observed in electronic systems since the 1980's, and the Chern-number measurement recently performed with ultracold bosonic atoms \cite{Aidelsburger:2014}.

\subsection{Fermions, the quantum Hall effect and the TKNN formula}\label{fermions}

The first situation that we consider is a 2D non-interacting polarized Fermi gas at zero temperature [Fig.~\ref{Goldman_fig2}]. Setting the Fermi energy $E_{\mathrm{F}}$ within a spectral gap naturally leads to a perfect filling of the bands $E_{\lambda}\!<\!E_{\mathrm{F}}$ located below the gap: the average number of particles in a state $u_{\lambda} (\bs{k})$, Eq.~(\ref{Goldman_eq4}), is exactly $\rho^{(\lambda)}\!=\!N^{(\lambda)}/N_{\mathrm{states}} \!=\!1$ for $E_{\lambda}\!<\! E_{\mathrm{F}}$. Setting the latter condition into Eq.~(\ref{Goldman_eq5}) yields
\begin{eqnarray}
&&v^x_{\mathrm{tot}}=  - \frac{F_y A_{\mathrm{syst}}}{h} \sum_{E_{\lambda}\!<\! E_{\mathrm{F}}}  \nu_{\mathrm{ch}}^{(\lambda)}  , \label{Goldman_eq7}
\end{eqnarray}
which indicates that the total velocity of the Fermi gas is directly related to the sum of Chern numbers associated with populated bands.

In systems presenting time-reversal symmetry, the Berry curvature satisfies $\Omega_{\lambda} (\bs{k})\!=\!-\Omega_{\lambda} (-\bs{k})$, in which case all the Chern numbers  $\nu_{\mathrm{ch}}^{(\lambda)}\!=\!0$, see Eq.~(\ref{Goldman_eq6}).  Hence, observing the transverse drift associated with $v_{\mathrm{tot}}^x$ requires a system without time-reversal symmetry, which is the case, e.g., in electronic systems subjected to magnetic fields \cite{Bernevig}. In these electronic setups, the transverse transport predicted by Eq. (\ref{Goldman_eq7}) is measured through the Hall conductivity $\sigma_{xy}$, which relates the electric field $E_y$ to the current density $j_x \!=\! e v_{\mathrm{tot}}^x/A_{\mathrm{syst}}$, where $e$ is the electron charge. Using these definitions together with Eq. (\ref{Goldman_eq7}), we recover the well-known TKNN formula \cite{Thouless1982} for the electric Hall conductivity
\begin{equation}
j_x=\sigma_{xy} E_y , \quad \sigma_{yx} =  \frac{e^2}{h}  \sum_{E_{\lambda}\!<\! E_{\mathrm{F}}}\nu_{\mathrm{ch}}^{(\lambda)} = - \sigma_{xy},\label{Goldman_eq8}
\end{equation}
where we introduced the electric field $E_y = F_y/e$ acting on the electrons. The TKNN formula (\ref{Goldman_eq8}) expresses the fact that transport measurements directly reveal the Chern numbers $\nu_{\mathrm{ch}}^{(\lambda)}$ in electronic systems, through the quantization of the Hall conductivity.  In particular, the quantized value $\sigma_{yx}=(e^2/h) \times$ (integer) remains constant as long as the Fermi energy stays in an open gap  \cite{Thouless1982,Kohmoto:1985}. Note that populating a Chern band with a Fermi gas is not limited to electronic systems, as it could also be performed by trapping fermionic atoms in an optical lattice subjected to artificial magnetic fields~\cite{Dauphin:2013}.

\subsection{Thermal Bose gas and the center-of-mass displacement}\label{thermal}

Let us now consider a radically different configuration: a thermal gas of non-interacting polarized bosons, whose temperature is large compared to the bandwidth  $W_{\mathrm{band}}$ of the lowest band $E_1$, but small compared to the spectral gap $\Delta$ above it [Fig.~\ref{Goldman_fig2}]. In this case, the average number of bosons in a state of the lowest band is homogeneous but density-dependent, $\rho^{(1)}\!=\!=\!N_{\mathrm{tot}}/N_{\mathrm{states}} \!\ne\!1$, and $\rho^{(\lambda)}=0$ for $\lambda>1$. The homogeneity of the band filling, which is a reasonable assumption when the lowest band presents a large flatness ratio $\Delta/W_{\mathrm{band}}\! \gg \!1$, can be tested in cold-atom experiments using band-mapping techniques  \cite{Bloch2008,Aidelsburger:2014}. Setting the particle filling condition into Eq.~(\ref{Goldman_eq5}) yields
\begin{eqnarray}
&&v^x_{\mathrm{tot}}=  - \frac{F_y A_{\mathrm{syst}}}{h} \,  \rho \, \nu_{\mathrm{ch}}^{(1)}  , \quad \rho=N_{\mathrm{tot}}/N_{\mathrm{states}} \ne 1 ,\label{Goldman_eq9}
\end{eqnarray}
which differs from the Fermi-gas result in Eq.~(\ref{Goldman_eq7}) in that only the lowest band contributes, but also through the additional density dependence. 

The Bose gases considered in cold-atom experiments are charge neutral. However, in analogy with the quantum Hall effect discussed in Section \ref{fermions}, one could consider a transport measurement relating the particle current density $j_x \!=\! v_{\mathrm{tot}}^x/A_{\mathrm{syst}} $ to the applied force $F_y$. In this case, using Eq.~(\ref{Goldman_eq9}), we find that the analogue of the Hall conductivity would read
\begin{equation}
j_x=\sigma_{xy} F_y , \quad \sigma_{xy} = -  (1/h) \,  \rho \, \nu_{\mathrm{ch}}^{(1)}.\label{Goldman_eq10}
\end{equation}
In contrast with the TKNN formula (\ref{Goldman_eq8}), the relation between the measured transport coefficient $\sigma_{xy}$ and the topological Chern number involves a density-dependent factor $\rho/h$. Thus, in such an experiment, identifying the Chern number of the lowest band would  require to combine simultaneous transport and density measurements, which constitutes a severe drawback of this approach.  To overcome this difficulty, one can consider another physical observable: the center-of-mass (CM) displacement of the gas. Indeed, using Eq.~(\ref{Goldman_eq9}), we find that the transverse velocity of the CM is given by 
\begin{eqnarray}
&&v^x_{\mathrm{CM}}=v^x_{\mathrm{tot}}/N_{\mathrm{tot}}=  - \frac{F_y A_{\mathrm{cell}}}{h} \,  \, \nu_{\mathrm{ch}}^{(1)} .\label{Goldman_eq11}
\end{eqnarray}
Since both the unit cell area $A_{\mathrm{cell}}$ and the strength of the applied force $F_y$ can be determined with precision, the transverse displacement of the CM $\Delta x_{\mathrm{CM}}(t)=v^x_{\mathrm{CM}} t$ offers a direct measure of the Chern number of the lowest band. This Chern-number-measurement was successfully implemented in Munich \cite{Aidelsburger:2014} in 2014, see Section \ref{Dalibard_Implementation}.

\section{Models: Harper-Hofstadter bands vs. Haldane-like bands}\label{Dalibard_models}

We now turn to the presentation of two lattice models that have been recently implemented with cold atoms setups and that lead to a topologically non-trivial lowest band. For the sake of completeness, we start our discussion with a short reminder of the well-known Landau levels, which characterize the quantum motion of a free particle in a uniform magnetic field. This Landau level structure, with its non trivial bulk topological features and the associated edge currents, is crucial to explain Quantum-Hall-type phenomena, both in the integer and the fractional cases.

\subsection{Landau levels}

The Landau level spectrum \cite{Cohen} emerges when one looks for the eigenstates of the Hamiltonian $\hat H= (\hat {\bs p}-q\bs A(\hat {\bs r}))^2/2m$, where $\hat {\bs p}$ (resp. $\hat {\bs r}$) denotes the canonical momentum (resp. position) operator for a particle of charge $q$ and mass $m$. The vector potential $\bs A(\bs r)$, defined up to a gauge transformation, satisfies $\bs \nabla \times \bs A=\bs B$, where $\bs B$ is uniform. In order to derive the spectrum of $\hat H$, we note that it can be written as $\hat H= (\hat \Pi_x^2+\hat \Pi_y^2)/2m$, where we introduced the kinetic momentum $\hat {\bs \Pi}=\hat {\bs p}-q\bs A(\hat {\bs r})$. The kinetic momentum operators satisfy the simple commutation relation $[\hat \Pi_x,\hat \Pi_y]=i \hbar q B$. The corresponding operator algebra is thus formally equivalent to that of a harmonic oscillator where $\hat H=(\hat X^2+\hat P^2)/2$ with $[\hat X,\hat P]=i$. We infer from this equivalence that the energy levels form an equidistant set $\hbar \omega_c(n+1/2)$, where $\omega_c=qB/m$ denotes the cyclotron frequency and $n$ is a nonnegative integer. Each Landau level has a macroscopic degeneracy $S/(2\pi \ell^2)$, where $S$ is the area of the sample and $\ell= (\hbar/qB)^{1/2}$ is the magnetic length. One thus obtains a band-like spectrum as for a particle in a periodic lattice, each band being infinitely narrow. Each Landau level has a Chern number equal to 1, so that the notion of anomalous velocity is also relevant here. Actually it has in this case a simple classical interpretation in terms of Hall current. It is indeed well known that when a particle in a magnetic field along $z$ is acted upon by an additional uniform force $F_y$, its motion consists in the combination of the circular cyclotron motion and a uniform translation motion at the constant velocity $v_x=F_y/(qB)$; the latter is nothing but the anomalous velocity described above. 

 \begin{figure}[t]
\begin{center}
\includegraphics{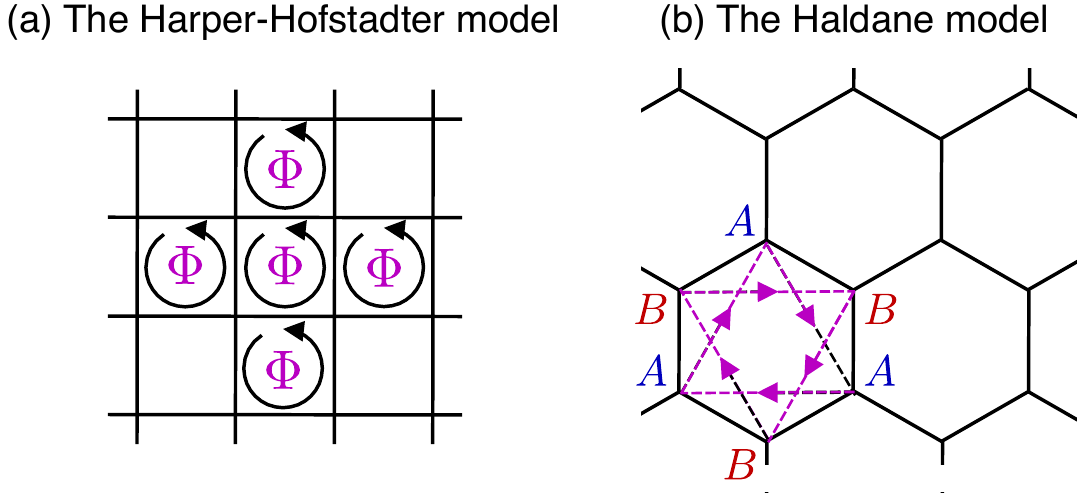}
\end{center}
\caption{Two lattice models leading to topologically non-trivial bands. (a) The Harper-Hofstadter model: a square lattice operating in the tight-binding regime is placed in a uniform field, with the flux $\Phi$ across each unit cell of the lattice \cite{Harper:1955,Hofstadter1976}. (b) The Haldane model: a periodic honeycomb lattice with nearest-neighbour couplings $A\leftrightarrow B$ (continuous lines) corresponding to a real tunneling matrix element, and next-to-nearest neighbour couplings $A\leftrightarrow A$ and $B\leftrightarrow B$ (dotted lines) corresponding to a non-real tunneling matrix element. The arrow directions encode the sign of the argument of this matrix element \cite{Haldane1988}. }
\label{Goldman_fig3}
\end{figure}

\subsection{Lattices and tight-binding models}\label{section:models}

We now turn to the Harper-Hofstadter problem \cite{Harper:1955,Hofstadter1976}, which is the transposition of the Landau problem to a discretized space. It models the motion of a charged particle in a square two-dimensional lattice of period $a$, normal to the uniform magnetic field (Fig.\ref{Goldman_fig3}(a)). This motion is described using the single band approximation, assuming that only nearest neighbour couplings are relevant (Hubbard model). The quantum state corresponding to the particle localized on the lattice site $\bs r=a(m,n)$ is denoted $|\bs r\rangle$, with $m,n$ integers. The Hamiltonian reads in zero magnetic field
\begin{equation}
\hat H_0=-J \sum_{\langle \bs r, \bs r'\rangle} \hat a_{\bs r'}^\dagger \hat a_{\bs r},
\label{Goldman_eq12}
\end{equation}
where $\hat a_{\bs r}$ annihilates a particle in the state $|\bs r\rangle$ and $J>0$ is the tunneling matrix element. The eigenstates of this Hamiltonian are the Bloch functions $|\psi_{\bs k}\rangle=\sum_{\bs r} e^{i \bs k\cdot \bs r} |\bs r\rangle$ with the energy $E(\bs k)=-2J\left[\cos(k_xa)+\cos(k_ya)\right]$, corresponding to an allowed band of width $8J$. 

The presence of the magnetic field is taken into account in the single-band approximation by assigning a complex value to the tunneling matrix elements between adjacent sites. The individual phases of these matrix elements have to be chosen such  that the total Aharonov--Bohm phase 
\cite{Aharonov-Bohm59PR} accumulated along a closed contour  is equal to $2\pi \Phi/\Phi_0$, where $\Phi$ is the flux of the magnetic field through the contour and $\Phi_0=h/q$ is the flux quantum  \cite{Hofstadter1976,GoldmanReview}. Because of gauge invariance, there are an infinite number of choices for individual phases matching this prescription. Here we  choose the Landau gauge, which amounts to taking  zero phase (a real matrix element) for  tunneling along the $y$ direction, and the $y$-dependent phase $2\pi \alpha n$ for the tunneling along the $x$ direction, where we introduced the dimensionless flux parameter $\alpha=\Phi/\Phi_0$. The Hamiltonian is now
\begin{equation}
\hat H_\alpha=-J \sum_{m,n} \left( \hat a_{m,n+1}^\dagger \hat a_{m,n} + e^{i 2\pi \alpha n} \hat a_{m+1,n}^\dagger \hat a_{m,n}\right) + \mbox{H.c.}
\label{Goldman_eq13}
\end{equation} 
which is periodic in $\alpha$ with period 1. The term $\exp (i 2\pi \alpha n)$, which captures the effect of the magnetic field, is generally referred to as the Peierls phase-factor \cite{Harper:1955,Hofstadter1976}.

The single-particle spectrum plotted as a function of $\alpha$ acquires a fractal structure known as the \emph{Hofstadter butterfly}. The calculation of this spectrum for an arbitrary value of $\alpha$ is a difficult mathematical task. If one restricts to rational values $\alpha=p/q$ ($p$ and $q$ relatively prime integers), the problem is simpler since one recovers a two-dimensional periodic problem, with a period $a$ along $x$ and $qa$ along $y$. The corresponding unit cell (the so-called \emph{magnetic cell}) now contains $q$ sites, and the energy band of width $8J$ for $\Phi=0$  split in $q$ non-overlapping subbands (two adjacent subbands may touch each other via Dirac points).  In general each subband has a nonzero Chern index, which can be obtained via the solution of a Diophantine equation \cite{Thouless1982}, whereas the sum of all Chern indices over the $q$ subbands is null.  In the particular case $\alpha=1/q$, the Chern index of the lowest band is 1, so that this band is topologically equivalent to the lowest Landau level found in the absence of a lattice.

In the Harper-Hofstadter problem, the presence of the uniform magnetic field breaks the translational symmetry of the initial lattice.  Haldane proposed in 1988 another model \cite{Haldane1988}, in which a non-trivial band topology could appear without any modification of the lattice unit cell nor breaking of the translational symmetry. The starting point is a graphene-like honeycomb lattice with nearest-neighbour couplings (Fig.~\ref{Goldman_fig3}(b)). The unit cell thus consists of two equivalent sites, denoted $A$ and $B$, with tunneling from sites $A$ (resp. $B$) to the three neighbouring sites $B$ (resp. $A$) with equal amplitude. At this stage the single-particle spectrum consists of two subbands,  touching each other in Dirac points. Haldane's crucial insight was to add non-real next-to-nearest-neighbour (NNN) couplings, \emph{i.e.} $A\to A$ and $B\to B$, which break time-reversal symmetry. By contrast to the Harper-Hofstadter model, these couplings are constant over the whole lattice, and could in principle be created by a staggered magnetic field with a zero-flux through the honeycomb unit cell. These additional couplings lift the initial degeneracy at the Dirac points, and the two subbands are now separated by a gap with non-zero opposite Chern indices, $+1$ and $-1$. With the lowest band filled with spinless non-interacting fermions, Haldane's model constitutes a prototype of a Chern insulator that goes beyond quantum Hall setups. It played a key role for the subsequent discovery of time-reversal invariant topological insulators, where spin-orbit coupling replaces the complex  next-to-nearest-neighbour tunnel coupling. The optical flux lattice concept that will be presented latter in this Chapter is also directly related to Haldane's model, since it provides topologically non-trivial energy bands for configurations where the atom-light interaction is periodic.

\section{Implementation: driving atoms into topological matter}\label{Dalibard_Implementation}

This Section describes several schemes realizing Chern bands in cold-atom systems. For the sake of clarity, this Section makes the distinction between methods based on a tight-binding approach [Section \ref{TB}] and those applicable in the continuum or weak-lattice regimes [Section \ref{continuum}]. A special emphasis is set upon recent theoretical works and experimental implementations, which led to unambiguous signatures of topological properties associated with non-zero Chern numbers in 2D optical lattices. 

\subsection{Schemes based on a tight-binding approach}\label{TB}

Non-interacting cold atoms moving in a deep optical lattice \cite{Bloch2008} are well described by the single-band tight-binding Hamiltonian in Eq. \eqref{Goldman_eq12}. Starting with this topologically-trivial tight-binding band, it appears that a natural way to generate Chern bands consists in implementing the Harper-Hofstadter or the Haldane model [Section \ref{section:models}], which can be achieved by controlling the tunneling matrix elements within the optical-lattice setup, see Eq. \eqref{Goldman_eq13}. Indeed, as was discussed in Section \ref{section:models}, Peierls phase factors can be associated with (local) magnetic fluxes, which break TRS and  potentially lead to Chern bands. We now review several methods by which Peierls phase factors and magnetic fluxes can be engineered using different aspects of cold-atom technology, see Fig. \ref{Goldman_fig4}.

\subsubsection{Using internal states of the atoms}

\begin{figure}[h!]
\begin{center}
\includegraphics[width=12.5cm]{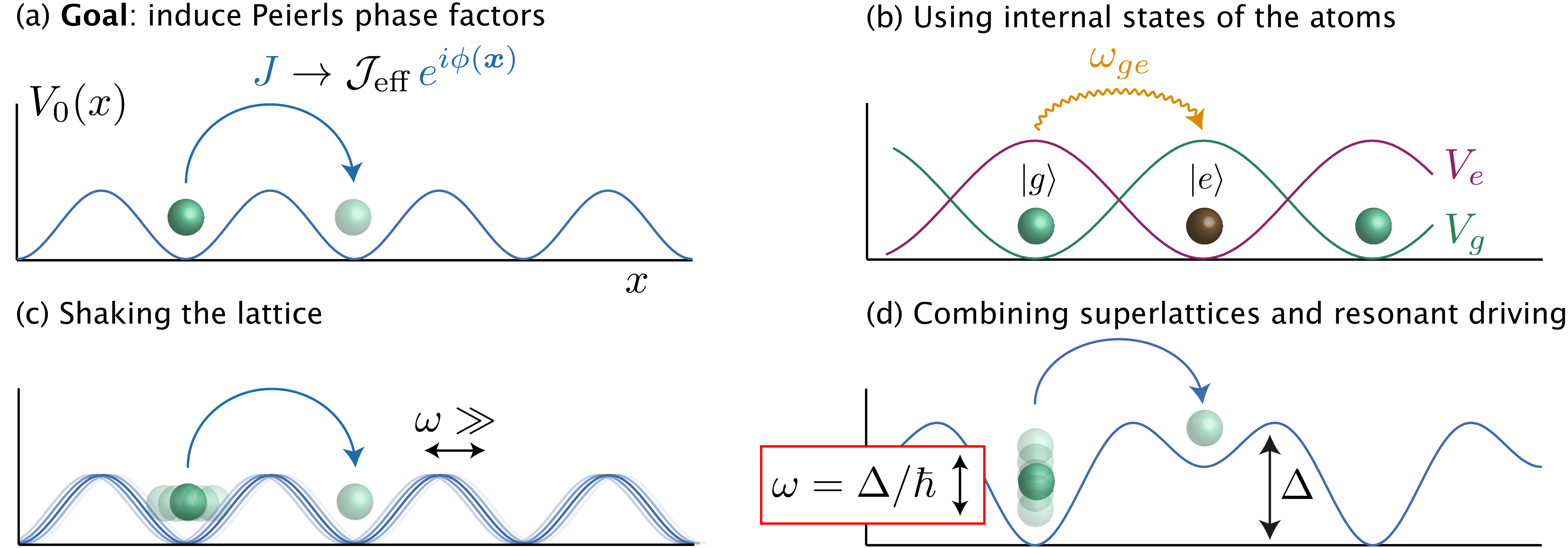} 
\caption{(a) Effective magnetic fluxes can be created in deep optical lattices by inducing complex tunneling matrix elements. (b) Method using state-dependent optical potentials $V_{g,e}$, trapping atoms in two internal states $\vert g \rangle$ and $\vert e \rangle$, combined with a resonant atom-light coupling with frequency $\omega_{ge}=(E_e-E_g)/\hbar$. (c) Shaking a lattice, with an arbitrarily large driving frequency $\omega$. (d) Method using a superlattice with energy offset $\Delta$, large compared to the bare hopping energy, combined with a resonant onsite energy modulation with frequency $\omega = \Delta/\hbar$.}
\label{Goldman_fig4}
\end{center}
\end{figure}

A natural way to induce complex tunneling matrix elements in optical lattices is to exploit the spatial-dependence of the optical phase in photon-assisted tunneling \cite{Jaksch2003,Gerbier:2010,Ruostekoski:2002fs,Mueller:2004hc}. Consider two internal states of an atom, denoted $\vert g \rangle$ and $\vert e \rangle$, respectively trapped in two independent (state-dependent) optical lattices $V_{g,e}$ along the $x$ direction, see Fig. \ref{Goldman_fig4}(b). For large enough lattice potentials, and in the absence of coupling between the two states, the hopping is completely inhibited. Adding a coherent coupling, with wave vector $\bs q$ and frequency $\omega_{ge}$ matching the energy difference between the internal states, effectively activates tunneling processes between the two sublattices. The tunneling matrix elements between two nearest-neighbouring lattice sites $\bs{r}_g$ and $\bs{r}_e$ are of the form $J (\bs{r}_g) = \mathcal{J}_{\text{eff}} \exp (i \bs q \cdot \bs{r}_g)$, where the amplitude $\mathcal{J}_{\text{eff}}$ is proportional to the coupling's Rabi frequency and to the overlap between Wannier functions defined at the two sites involved in the process \cite{Jaksch2003,Gerbier:2010,GoldmanReview,Dalibard2011}. Hence, space-dependent Peierls phase-factors, and the corresponding magnetic fluxes penetrating lattice plaquettes, can be controlled by tuning the coupling's wave vector $\bs q$. This method can be applied to square optical lattices, leading to an effective Harper-Hofstadter Hamiltonian \eqref{Goldman_eq13}, but also to triangular/honeycomb lattices in view of realizing the Haldane model \cite{Alba:2011,Goldman:2013,Anisimovas:2014ga}. The scheme is general and can be applied to diverse atom-light configurations: for instance, it may involve two hyperfine states in the ground-state manifold coupled through a (two-photon) Raman coupling \cite{Jaksch2003}, or a one-photon coupling between a ground state and a long-lived excited state \cite{Gerbier:2010}. This method has not yet been implemented experimentally. 

The scheme described above concerns the coupling between atoms living on a 2D optical lattice. However, it is intimately related to the concept of ``synthetic magnetic fields in synthetic dimensions" \cite{Celi:2013}, which has been implemented at LENS \cite{Mancini:2015} and NIST \cite{Stuhl:2015} in 2015. Here, atoms in $M$ internal states live on a standard 1D optical lattice, characterized by real-valued tunneling matrix elements. Atom-light coupling then drives on-site transitions between the $M$ internal states, such that each transition $m \rightarrow m+1$ is accompanied with a space-dependent phase $\bs q \cdot \bs{r}$ (i.e. a momentum transfer). Interpreting these internal-states transitions as hopping processes along a synthetic (internal-state) dimension, which is spanned by $m=1, \dots, M$, this system is found to be equivalent to the effective Harper-Hofstadter system described above. The propagation of chiral edge-states, an unambiguous signature of synthetic magnetic fluxes in 2D lattices, has been experimentally measured in these experiments \cite{Mancini:2015,Stuhl:2015} .

\subsubsection{Superlattices and resonant time-modulations}

Complex tunneling matrix elements can also be engineered without considering the internal structures of the atoms. In direct analogy with the scheme described above, this can be achieved by imposing an energy offset between neighboring sites of an optical lattice and modulating the system with a resonant time-periodic modulation of the on-site energies, see Fig. \ref{Goldman_fig4}(d). In this case, the Peierls phase-factors are directly related to the phase of the modulation.  To be explicit, let us consider a minimal lattice system: two lattice sites ($a$ and $b$), separated in energy by a large offset $\Delta$, and subjected to a resonant modulation acting on the lower site ($a$) only. We write the tight-binding Hamiltonian in the  form
\begin{equation}
\hat H (t) = -J \left ( \hat a^{\dagger} \hat b + \hat b^{\dagger} \hat a  \right ) + \Delta \, \hat b^{\dagger} \hat b  + \kappa \cos (\omega t + \phi) \hat a^{\dagger} \hat a ,\label{Goldman_eq14}
\end{equation}
where $J $ is the bare hopping amplitude between the sites. In the large-frequency regime $\Delta=\hbar \omega \gg J$, the long-time dynamics resulting from the time-dependent Hamiltonian $\hat H (t)$ is found to be captured by a time-independent (effective) Hamiltonian, which in this case, takes the simple form \cite{Eckardt:2007it,Lim2010,Bermudez:2011prl,Hauke:2012,Goldman:2014vs}
\begin{equation}
\hat H_{\text{eff}} = -J \mathcal{J}_1 (\kappa / \hbar \omega) \,  \exp (i \phi) \,  \hat a^{\dagger} \hat b+ \text{h.c.} ,\label{Goldman_eq15}
\end{equation}
where $\mathcal{J}_1$ denotes the Bessel function of the first kind. According to Eq. \eqref{Goldman_eq15}, the tunneling between the sites is effectively restored and the corresponding tunneling matrix elements include a Peierls phase-factor related to the phase of the modulation $\phi$, which can be made space-dependent. The result in Eq. \eqref{Goldman_eq15} constitutes the building blocks for the generation of magnetic fluxes in time-modulated optical lattices, as implemented in Munich \cite{Aidelsburger:2011,Aidelsburger:2013,Aidelsburger:2014} and at MIT \cite{Ketterle:2013}. The following of the Section describes these schemes in more detail. 

In order to treat time-modulated optical lattices, we first introduce a set of useful equations, which offer a powerful theoretical framework to analyse the physics of time-dependent problems. Let us consider a general Hamiltonian of the form $\hat H(t) = \hat H_0 + \hat V(t)$, where $\hat V(t+T)=\hat V(t)$ and where $T=2 \pi /\omega$ is the period of the driving. In the following, the period $T$ is considered to be small compared to all characteristic time-scales \cite{Goldman:2014PRX}.  We are interested in describing the dynamics of a initial state $\psi (t_0)$ under the driving, which is assumed to start at time $t_0$. Formally, the state at a given time $\psi (t)$ is obtained by acting on the initial state $\psi (t_0)$ with the time-evolution operator $\hat U (t ; t_0)$, which can be partitioned as \cite{Rahav:2003,Goldman:2014PRX}
\begin{equation}
\hat U (t ; t_0) = \exp [-i \hat K (t)] \exp [-i (t-t_0) \hat H_{\text{eff}}/\hbar] \exp [i \hat K (t_0)],\label{Goldman_eq16}
\end{equation}
where $\hat H_{\text{eff}}$ is a time-independent (effective) Hamiltonian describing the long-time dynamics, and where the ``kick" operator $\hat K (t)$ captures the micro-motion and the effects related to the initial phase of the modulation. The dynamics is thus completely captured by the operators $\hat H_{\text{eff}}$ and $\hat K (t)$, which, according to Refs. \cite{Rahav:2003,Goldman:2014PRX,Goldman:2014vs}, can be calculated systematically through a perturbative expansion in powers of $1/\omega$. Considering a single-harmonic modulation, $\hat V (t) = \hat V \exp (i \omega t) + \text{h.c.}$, these operators are approximatively given by \cite{Goldman:2014PRX}
\begin{align}
&\hat H_{\text{eff}}= \hat H_0 + (1/\hbar \omega) [\hat V, \hat V^{\dagger}] + 1/( \sqrt{2}\hbar \omega)^2 \left ( [[\hat V, \hat H_0],\hat V^{\dagger}]  + \text{h.c.}  \right ) + \mathcal O (1/\omega^3), \notag \\
&\hat K (t)=(1/i \hbar \omega) \left ( \hat V \exp (i \omega t) - \text{h.c.} \right ) + \mathcal O (1/\omega^2).\label{Goldman_eq17}
\end{align}
In some cases, the infinite series in Eq. \eqref{Goldman_eq17} can be (partially or totally) resummed; this allows  convergence of the expansion even in the strong-driving regime \cite{Goldman:2014PRX,Goldman:2014vs,Bukov:2014vd}. The equations \eqref{Goldman_eq16}-\eqref{Goldman_eq17} offer a systematic way to build the time-evolution operator, which can be directly applied to arbitrarily complicated systems. 

Having defined a theoretical framework to treat time-dependent problems, we now apply it to time-modulated optical superlattices. The Hamiltonian is taken in the form $\hat H(t) = \hat H_0 + \hat V(t)$, where the static part consists of a 2D tight-binding Hamiltonian with an additional superlattice potential directed along the $x$ direction
\begin{equation}
\hat H_0= - \sum_{m,n} \left ( J_x  \hat a_{m+1,n}^{\dagger} \hat a_{m,n} + J_y \hat a_{m,n+1}^{\dagger} \hat a_{m,n} + \text{h.c.} \right )+  \Delta \sum_{m,n} s(m) \hat a_{m,n}^{\dagger} \hat a_{m,n},\label{Goldman_eq18}
\end{equation}
where $\hat a_{m,n}$ creates an atom at lattice site $\bs r= a(m,n)$, $a$ is the spacing and $(m,n)$ are integers. The superlattice function $s(m)$ is assumed to create energy offsets  $\Delta \gg J_{x}$ between all neighboring sites, i.e. $s(m+1)- s(m)= \pm 1$ for all $m$. The tunneling is then restored by applying an on-site time-periodic modulation of the form 
\begin{equation}
\hat V (t) = \hat V \exp (i \omega t) + \text{h.c.}, \quad \hat V = \kappa \sum_{m,n} v(m,n) \hat a_{m,n}^{\dagger} \hat a_{m,n} .\label{Goldman_eq19}
\end{equation}
Here, the complex numbers $v(m,n)$ capture the space-dependence of the modulation, which turns out to be crucial, and we impose the resonance condition $\omega = \Delta/\hbar$. The long-time dynamics of the modulated 2D lattice [Eqs. \eqref{Goldman_eq18}-\eqref{Goldman_eq19}] is captured by the effective Hamiltonian $\hat H_{\text{eff}}$ in Eq. \eqref{Goldman_eq17}, which after partial resummation of the series yields a Harper-Hofstadter-like Hamiltonian \cite{Goldman:2014vs}
\begin{align}
&\hat H_{\text{eff}} \!=\! \sum_{m,n} \mathcal{J}_x (m,n)e^{\pm i \theta (m,n)}  \hat a_{m+1,n}^{\dagger} \hat a_{m,n} \!+\! \mathcal{J}_y (m,n) \hat a_{m,n+1}^{\dagger} \hat a_{m,n} \!+\! \text{h.c.} , \label{Goldman_eq20}
\end{align}
where $\theta (m,n)=\text{arg} [v(m+1,n) - v(m,n)]$, and where the sign $\pm$ depends on whether the hopping $m \rightarrow m+1$ starts from a low- or a high-energy site of the superlattice potential $s(m)$. Hence, the Peierls phase-factors in the effective Hamiltonian (and the corresponding magnetic fluxes per plaquette) depend on the time-modulation $\hat V(t)$, through $v(m,n)$ [Eq. \eqref{Goldman_eq19}], but also on the static superlattice potential $s(m)$ [Eq. \eqref{Goldman_eq18}]. The effective hopping amplitudes $\mathcal{J}_{x,y}$ are \emph{a priori} space-dependent (see \cite{Kolovsky:2011} and the discussion in \cite{Creffield:2013gp}); they are explicitly given by 
\begin{equation}
\mathcal{J}_x (m,n)\!=\! J_x \mathcal{J}_1 [K_0 \vert \delta_x v(m,n) \vert], \, \mathcal{J}_y (m,n)\!=\! J_y \mathcal{J}_0 [K_0 \vert \delta_y v(m,n) \vert], \notag
\end{equation}
where $K_0=\!2 \kappa/\hbar \omega$ and $\delta_{x,y}v$ denote finite-difference operations along the $x$ and $y$ directions, e.g. $\delta_{x} v(m,n)\!=\! v(m+1,n) \!-\! v(m,n)$. The kick operator can be obtained along the same line; its impact on physical observables (e.g. momentum distributions) was discussed in Ref. \cite{Goldman:2014vs}. 

Based on the result \eqref{Goldman_eq20}, two different features of the system should be designed so as to eventually realize the Harper-Hofstadter Hamiltonian \eqref{Goldman_eq13}: the superlattice function $s(m)$ and the time-modulation function $v(m,n)$. The main difficulty to achieve the uniform-magnetic flux configuration is to handle the sign of the phase $\pm \theta (m,n)$, which can typically lead to staggered flux patterns in arbitrary superlattices \cite{Aidelsburger:2011}. To solve this issue, the Munich \cite{Aidelsburger:2013} and MIT \cite{Ketterle:2013} teams first combined  a Wannier-Stark ladder potential generated by a magnetic-field gradient, i.e. $s(m)=m$, with a simple moving potential, $v(m,n)= \exp (i \bs q \cdot \bs r)$, created by a single pair of laser beams. In this configuration, a uniform flux $2 \pi \alpha=q_y a$, together with homogeneous hopping amplitudes $\mathcal{J}_{x,y}$, were achieved. By setting $q_y = \pi /2 a$, the Munich team obtained a topological band structure, whose lowest band had a Chern number $\nu_1 = 1$. However, this setup was found to be unsuitable for the Chern-number measurement described in Section \ref{thermal}, due to instabilities arising from the linear gradient $s(m)=m$.  In order to achieve uniform flux with an all-optical superlattice \cite{Baur:2014ux,Aidelsburger:2014}, the Munich team then developed a novel setup, based on a two-site superlattice, $s(m)=(-1)^m$, combined with a more sophisticated time-modulation of the lattice\textsuperscript{3}\footnotetext[3]{Combining a two-site superlattice to a simple moving potential with $v(m,n)= \exp (i \bs q \cdot \bs r)$ leads to a staggered flux configuration, which is associated with zero Chern numbers  \cite{Aidelsburger:2011}. The time-modulation used in  \cite{Aidelsburger:2014} allowed to rectify the flux, by individually addressing successive links with two independent pairs of lasers.}. The latter was induced by two pairs of laser beams, so as to generate a uniform flux over the entire lattice \cite{Aidelsburger:2014,Goldman:2014vs}, hence providing a stable platform to measure the Chern number, as we now describe.   

The Chern-number measurement \cite{Aidelsburger:2014} was achieved by loading bosonic atoms into the lowest band of the Harper-Hofstadter spectrum for a synthetic magnetic flux $\alpha=1/4$. The transverse drift of the cloud [Eq. \eqref{Goldman_eq11}] was then detected \emph{in-situ}, as a response to a weak optical gradient applied along the $y$ direction. For short times, the transverse motion of the cloud was found to be linear, in agreement with the prediction in Eq. \eqref{Goldman_eq11}. For longer-times, heating processes, which were found to be independent of the applied force, promoted atoms to higher bands and lead to a saturation of the cloud's transverse drift. A careful analysis taking into account the dynamical repopulations of the bands revealed the expected Chern number with a precision at the 1 $\%$ level. This Chern-number measurement was a direct probe for the topological order associated with the effective bulk energy bands, which resulted from the modulated optical superlattice.

\subsubsection{Off-resonant shaken optical lattices}

Before ending this Section on tight-binding realizations of Chern bands, let us briefly describe another promising strategy for creating and observing topological properties in optical-lattice systems. Similarly to the schemes discussed above, this method is based on time-periodic driving of the lattice, however, the driving frequency is now considered to be \emph{off-resonant} with respect to any energy separations in the problem, see Fig. \ref{Goldman_fig4}(c). Let us compare this idea with the resonant-modulation approach, by re-writing the two-site minimal model in Eq. \eqref{Goldman_eq14} as
\begin{equation}
\hat H (t) = -J \left ( \hat a^{\dagger} \hat b + \hat b^{\dagger} \hat a  \right ) + \kappa \cos (\omega t + \phi) \hat a^{\dagger} \hat a ,\label{Goldman_eq21}
\end{equation}
and by computing the corresponding effective Hamiltonian \eqref{Goldman_eq17} \cite{Arimondo2012,Goldman:2014PRX}
\begin{equation}
\hat H_{\text{eff}} = -J \mathcal{J}_0 (\kappa/\hbar \omega) \left ( \hat a^{\dagger} \hat b + \hat b^{\dagger} \hat a  \right ).\label{Goldman_eq22}
\end{equation}
Contrary to the resonant-modulation case [Eq. \eqref{Goldman_eq15}], the driving phase $\phi$ does not contribute to any Peierls phase-factor. In fact, a non-sinusoidal driving is required to generate effective complex tunneling matrix elements. This was demonstrated in Hamburg in 2012, where a Peierls phase factor was shown to appear in a shaken 1D optical lattice \cite{Struck2012}. The shaking strategy was also applied in 2D triangular lattices by the same team, in view of studying frustrated magnetism in optical lattices \cite{Struck2011,Struck:2013}. While these setups indeed produce local magnetic fluxes, their corresponding band structure is associated with zero Chern numbers. 

In 2014, Jotzu \emph{et al.} achieved the shaking of a 2D honeycomb lattice in a circular manner \cite{Jotzu:2014}. Similarly to the case of rotating traps [Section \ref{continuum}], this fast circular motion induced a chirality within the system, which formally breaks time-reversal symmetry. In fact, the effective Hamiltonian was shown to be equivalent to the Haldane model presented in Section \ref{section:models}: the circular shaking effectively induces complex NNN tunneling matrix elements. This opens a bulk gap in the honeycomb-lattice (Dirac) spectrum, and it generates Chern bands. The Berry curvature of the bands was probed in the system \cite{Jotzu:2014}, through the observation of an anomalous velocity.

\subsection{Schemes in the continuum}\label{continuum}

A natural route to the formation of topological bands with non-zero Chern number is to simulate the orbital effect of a uniform magnetic field. The atom should move continuously through space, subject only to this uniform field.  This leads to the Landau level spectrum: highly degenerate bands, each with unit Chern number.  There exist several ways to achieve this goal for cold atoms, at least to a good approximation. (Deviations can arise from non-uniformity of the field and/or the presence of additional potentials.)

\subsubsection{Rotation}

One very direct approach is to cause the atomic cloud to rotate, for example by applying a rotating deformation and bringing the system to equilibrium in a frame of reference rotating at angular frequency $\bs{\Omega}$ \cite{Cooper2008,fetter}. In this rotating frame an atom of mass $m$ moving with velocity $\bs{v}$ experiences a Coriolis force $\bs{F}_{\rm C} =  2m \bs{v} \times \bs{\Omega}$. This plays the same role as the Lorentz force on a charged particle in a uniform magnetic field, $\bs{  F}_{\rm L} = q\bs{v} \times \bs{B}$. Equating coefficients,
 $ q\bs{B} = 2m \bs{\Omega}$, one deduces an effective flux density $n_\phi \equiv {q|\bs{B}|}/{h} = {2m|\bs{\Omega}|}/{h}$. For a  Bose-Einstein condensate, this manifests itself as a lattice of quantized vortex lines of areal density $n_\phi$. Such vortex lattices have been observed, and their properties studied in
detail, in experiments dating back over 15 years \cite{MadisonCWD00,abos01,hodby:010405,coddington:063607}.  For a circularly symmetric harmonic confinement of frequency $\omega_0$,  mechanical stability under the centrifugal force sets an upper limit on rotation rate $|\bs{\Omega}| \leq
\omega_0$. At the point of balance,  $|\bs{\Omega}| = \omega_0$,  the spectrum for the 2D motion  is
exactly that of Landau levels.  This condition $|\bs{\Omega}| \simeq \omega_0$ has been reached experimentally to within $1\%$ \cite{SchweikhardCEMC92}. This rotation technique can also be implemented with optical lattices, see e.g.   \cite{Tung:2006b,Hemmerich:2007,Williams:2010}.

\subsubsection{Raman dressing}

\label{subsec:raman}

An important feature of the physics of cold gases is that  atoms (or molecules) can be prepared in different internal states (spin states or electronic excited states) which can have long lifetimes. It is then possible to use
optical fields to drive atoms into coherent superpositions of these internal states. Optical dressing of internal degrees of freedom provides  very powerful ways to generate artificial gauge fields and to form optical lattices with topological bands.

The physics underlying the use of internal states to generate artificial gauge fields can be understood in terms of a Berry connection in real space \cite{Berry:1984,Dum:1996}. Consider an atom with two internal states, $\alpha =1,2$. We denote the eigenstates at position $\bs{r}$ for these internal states by $|1\rangle_{\bs{r}}$ and $|2\rangle_{\bs{r}}$. In the presence of (optical) fields that couple the internal states, and in the rotating wave approximation, the local energy eigenstates are dressed states $|n\rangle_{\bs{r}} = a_{n,1}(\bs{r}) |1\rangle_{\bs{r}} + a_{n,2} (\bs{r})|2\rangle_{\bs{r}}$. The coefficients $a_{n,\alpha}(\bs{r})$ are determined by the fields at
position $\bs{r}$, so vary in space. Hence, the adiabatic motion of an atom in the dressed state $|n\rangle_{\bs{r}}$ is associated with a Berry connection $\bs{A}_n(\bs{r}) = i \hbar \langle n| \bs{\nabla} | n\rangle_{\bs{r}} = i\hbar \left(a_{n,1}^* \bs{\nabla}a_{n,1} + a_{n,2}^* \bs{\nabla}a_{n,2} \right)$. This plays the role of a vector potential coupling to the motion of the atom.  By careful choice of the spatial dependence of the fields, one can arrange that not only $\bs{A}_n$ but also $\bs{\nabla} \times \bs{A}_n$ is nonzero (for example for the lowest energy dressed state), such that the particle experiences a nonzero artificial
magnetic field \cite{Dalibard2011}. This method of generating artificial gauge fields has been demonstrated
in experiments at NIST \cite{spielmanfield} using Raman coupling of
spin states of rubidium, see the contribution by I. B. Spielman \emph{et al.}.  

For the method of Ref.~\cite{spielmanfield}
the vector potential is limited to $|\bs{A}| \lesssim h/\lambda_{\rm r}$ where $\lambda_{\rm r}$ is the optical wavelength.  By
Stokes' theorem, the total number of flux quanta through a circular
atomic cloud of radius $R$ is then $N_\phi \equiv \frac{1}{\hbar} \int
\bs{\nabla}\times \bs{A} \; d^2 r = \frac{1}{\hbar} \oint \bs{A}
\cdot d\bs{l} \lesssim R/\lambda_{\rm r}$.  Hence, the flux density $n_\phi
\equiv N_\phi/(\pi R^2)$ is limited to $n_\phi \lesssim 1/(R\lambda_{\rm r})$.
In Ref.~\cite{Cooper2011a} it was shown that periodic optical dressing
of internal states can allow the formation of  ``optical flux
lattices", with very much increased flux densities. These lattices lead to flux densities $n_\phi \sim
1/\lambda_{\rm r}^2$, larger by a factor of $R/\lambda_{\rm r}$ over previous proposals \cite{Dalibard2011}, which is
$ R/\lambda_{\rm r} \sim 50-100$ for typical experimental settings.

\subsubsection{Optical flux lattices}

\label{subsec:ofl}

The original  motivation to consider periodic optical lattices coupling internal
states\cite{Cooper2011a} was to find a way in which to increase the
strength of the artificial magnetic fields to $n_\phi\sim 1/\lambda_{\rm r}^2$.  Here we shall not focus on the real-space magnetic field. Instead, we concentrate on the band structure, defined in reciprocal space. We shall
show that two-dimensional optical lattices involving the coupling of internal states provide a very powerful way to
 generate topological (Chern) bands \cite{coopermoessner}.  We shall return at the end to comment on the connection to real-space magnetic field.

\begin{figure}
\includegraphics[width=1.0\columnwidth]{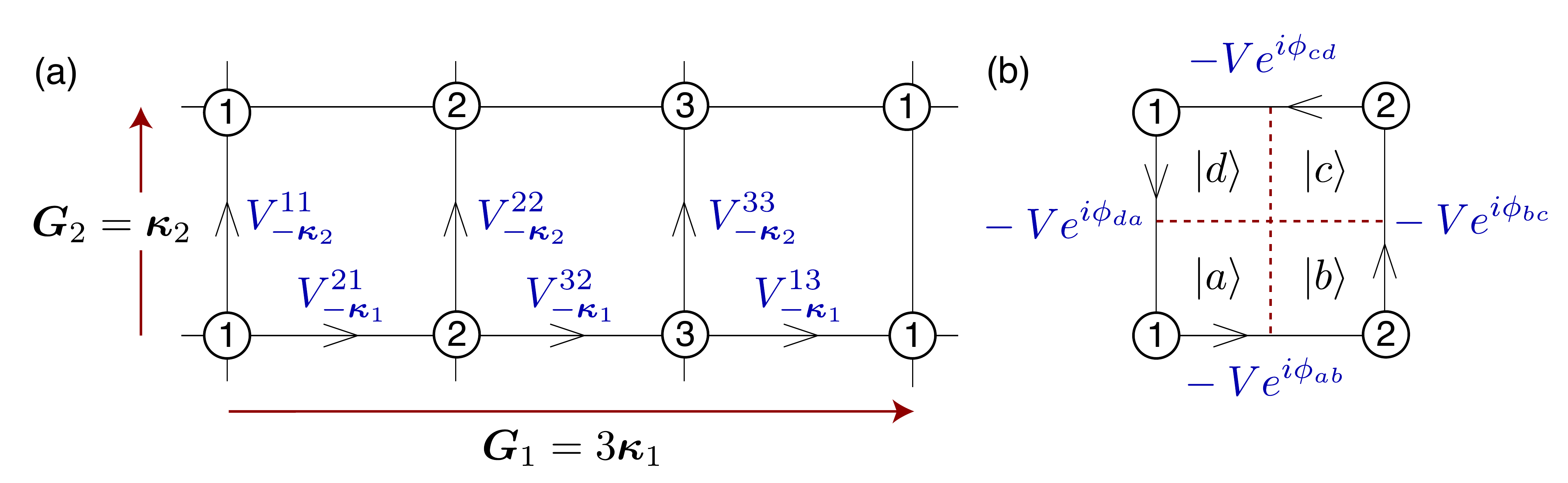}
\caption{(a) Example of the reciprocal space representation of an
  optical lattice involving couplings of $N_{\rm s} =3$ internal
  states. The sites denote the different values of $(\alpha, \bs{
    G})$, and the bonds the couplings $V^{\alpha \alpha'}_{\bs{
      g}_{\alpha' }+ \bs{G}' -\bs{g}_{\alpha} - \bs{G}}$. The
  offset momenta are $\bs{g}_1 = 0$, $\bs{g}_2 =\bs{
    \kappa}_1$, $\bs{g}_3 =2\bs{
    \kappa}_1$,  and the reciprocal lattice has basis vectors $\bs{
    G}_1= 3 \bs{\kappa}_1$, $\bs{G}_2 = \bs{\kappa}_2$.  (b) A
  representation of one plaquette of the reciprocal
  space lattice. For vanishing coupling $V=0$, the lowest energy state is one
  of the plane-wave states ($\{|a\rangle, |b\rangle, |c\rangle,
  |d\rangle\}$) defined in the text and is degenerate for
  quasi-momenta along the dashed lines. For nonzero $V$,
  gaps open along these lines. To first order in $V$, the integral of
  the Berry curvature of the lowest band over this plaquette is 
  by Eqn~(\protect\ref{Goldman_eq25}).}
\label{Goldman_fig5}
\end{figure}

Consider an atom with $N_{\rm s}$ long-lived internal states. We denote the (plane-wave) state of an atom of internal state $\alpha =1 \ldots N_{\rm s}$ and momentum $\hbar\bs{q}$ by $|\alpha, \bs{q}\rangle$. The coupling of the atom to optical fields will lead to processes which change the internal state and/or the momentum of the atom. We denote the optical coupling for  $|\alpha, \bs{q}\rangle \to |\alpha', \bs{q}'\rangle$ by $V^{\alpha'\alpha}_{\bs{q}'-\bs{q}}$. We consider the case of periodic lattices, for which the momentum transfers of the set of all such couplings, $V^{\alpha'\alpha}_{\bs{\kappa}}$, are commensurate.  The wave vector of 
any component $\alpha$ is then only conserved up to the addition of  reciprocal lattice vectors $\bs{G}$. 
 Similarly, by Bloch's theorem, the energy eigenstates can be assigned a band index $n$ and a quasi-momentum $\hbar \bs{k}$, and decomposed as
$|\psi^{n \bs{k}}\rangle = \sum_{ \alpha,\bs{G}} c^{n \bs{k}}_{\alpha,\bs{ G}}|\alpha, \bs{ k}-\bs{g}_\alpha - \bs{G}\rangle$ where $\bs{g}_\alpha$ accounts for possible momentum offsets of the different internal states\textsuperscript{4}\footnotetext[4]{This labeling of Bloch states takes advantage of an enhanced spatial symmetry of the system under combined translations and internal-state gauge-changes.
For the model illustrated in Fig.~\ref{Goldman_fig5}(a), the reciprocal lattice vector $\bs{G}_1$ is linked to a symmetry under the spatial translation by $\frac{1}{3}\frac{2\pi}{|\bs{\kappa}_1|}\hat{\bf \kappa}_1$ combined with the unitary transformation $\left[ |1\rangle\langle 1| + e^{i 2\pi/3} |2\rangle \langle 2| + e^{i 4\pi/3} |3\rangle \langle 3|\right]$.}. The band energies $E_n(\bs{k})$ follow from 
\begin{equation}
E_n(\bs{k}) c^{n\bs{k}}_{\alpha \bs{G}}  = \epsilon_{ \bs{k}-\bs{g}_\alpha - \bs{G}}  \; c^{n\bs{k}}_{\alpha\bs{G}}  + \sum_{\alpha',\bs{G}'} 
V^{\alpha \alpha'}_{
\bs{g}_{\alpha' }+ \bs{G}'
-\bs{g}_{\alpha} - \bs{G}}
\;c^{n\bs{k}}_{\alpha'\bs{G}'} 
\label{Goldman_eq23}
\end{equation}
where $\epsilon_{\bs{q}} \equiv \hbar^2|\bs{q}|^2/2m$ is the kinetic energy for an atom of momentum $\hbar \bs{q}$.

The structure of the couplings in 
Eqn (\ref{Goldman_eq23}) can be conveniently represented by a lattice in reciprocal space. An example for $N_{\rm s} = 3$  is shown in Fig.~\ref{Goldman_fig5}: the sites denote the different values of $(\alpha,  \bs{G})$; the bonds represent the couplings $V^{\alpha'\alpha}_{\bs{\kappa}}$. (We provide here no experimental implementation of this lattice, but note that it can be generated by a variant of the triangular lattice in Ref.~\cite{cooperdalibard2013}.)

It is instructive to construct the lowest energy band
using this picture in the ``nearly free particle" limit of small
$|V^{\alpha'\alpha}_{\bs{\kappa}}|$. Consider first vanishing coupling $|V^{\alpha'\alpha}_{\bs{\kappa}}| = 0$. From
Eqn~(\ref{Goldman_eq23}), at a given quasi-momentum $\hbar \bs{k}$, the
lowest energy state is that state with smallest $|\bs{k} - \bs{
  g}_\alpha - \bs{G}|^2$: that is, it is the plane wave state
$|\alpha, \bs{k}-\bs{g}_\alpha - \bs{G}\rangle $ associated with the lattice point
$(\alpha,\bs{G})$  closest to $\bs{k}$. This
divides each square plaquette into four regions, shown in
Fig.~\ref{Goldman_fig5}(b) for the leftmost plaquette of
Fig.~\ref{Goldman_fig5}(a), with the lowest energy state in these regions  being $|a\rangle = |1,
\bs{k}-\bs{g}_1\rangle$, $|b\rangle = |2, \bs{k}-\bs{g}_2\rangle$,
$|c\rangle = |2, \bs{k}-\bs{g}_2-\bs{G}_2\rangle$, $|d\rangle = |1,
\bs{k}-\bs{g}_1-\bs{G}_2\rangle$.  The dashed lines in
Fig.~\ref{Goldman_fig5}(b) show the locations where
 the
lowest energy state is degenerate for $V=0$: two-fold degenerate on each dashed
line and four-fold degenerate at the point  $\bs{k}^*$ where the dashed
lines cross. 

For $|V^{\alpha'\alpha}_{\bs{\kappa}}| \neq 0$, the degeneracies along the
dashed lines are lifted. We describe the behaviour in the case where
the couplings around this plaquette are $-Ve^{i\phi}$ with $\phi = \phi_{ab,bc,cd,da}$ as
illustrated in Fig.~\ref{Goldman_fig5}(b).  Working to lowest order in $V$ involves
first order (degenerate) perturbation. At the point of four-fold degeneracy $\bs{k}^*$ this leads to the
 Hamiltonian 
\begin{equation}
\left(\begin{array}{cccc}
 \epsilon_{\bs{k}^*}
 &   -V e^{-i\phi_{ab}} &  0
&   -V e^{i\phi_{da}}\\
 -V  e^{i\phi_{ab}} & \epsilon_{\bs{k}^*}
 & 
  -V e^{-i\phi_{bc}} & 0\\
 0 &  -V  e^{i\phi_{bc}} &  \epsilon_{\bs{k}^*} &  -V  e^{-i\phi_{cd}}\\
  -V e^{-i\phi_{da}}  & 0 &  -V  e^{i\phi_{cd}} & \epsilon_{\bs{k}^*}
\end{array}
\right)
\end{equation}
The spectrum of this 
Hamiltonian depends on the gauge invariant phase 
\begin{equation}
\Phi_{\rm p} \equiv  \phi_{ab}+\phi_{bc}+\phi_{cd}+\phi_{da}
\label{Goldman_eq25}
\end{equation} 
For $\Phi_{\rm p}\neq \pi$ the
lowest energy state is non-degenerate at $\bs{k}^*$, and indeed for
all $\bs{k}$ within this plaquette. The lowest band therefore
has a well-defined Berry curvature, with no singularities, for $\Phi_{\rm p}\neq \pi$. One can
then compute the integral of the Berry curvature over the plaquette in
terms of the line integral of the Berry connection around the four
sides of the plaquette. Still working to first order in $V$, the
adiabatic transfer along each of these sides just involves the
coupling of a pair of states (e.g. $|a\rangle$ and $|b\rangle$
coupled by $-Ve^{i\phi_{ab}}$). It is straightforward to show that the
total integral of the Berry connection around the plaquette is just
$\Phi_{\rm p}$. Thus, to first order in $V$, the integral of the Berry
curvature over the plaquette is  precisely $\Phi_{\rm p}$ for $\Phi_{\rm p}\neq \pi$. For $\Phi_{\rm p} =\pi$ the lowest two bands touch at a Dirac point at $\bs{k} = \bs{k}^*$,
and the Berry curvature of the lowest band is ill-defined.

This result provides a very powerful prescription by which to design
optical couplings that generate Chern bands.  By choosing the phases
of the couplings $V^{\alpha'\alpha}_\kappa$ around each plaquette in
the reciprocal space lattice, one can specify (the integral of) the
Berry curvature of the lowest energy band over each of these plaquettes. If the
sum over all plaquettes in the first Brillouin zone
 $\Phi_{\rm tot} =
\sum_{{\rm p}\in{\rm FBZ}} \Phi_{\rm p}$ is nonzero, and all $\Phi_{\rm p} \neq \pi$, then the lowest
energy band will have a Chern number of ${\cal C} = \Phi_{\rm
  tot}/(2\pi)$.  Thus the net number of ``flux quanta'' through this reciprocal space lattice sets the Chern number of the 
lowest energy band.
For example, if the couplings of 
Fig.~\ref{Goldman_fig5}(a) are chosen 
\begin{eqnarray}
V^{21}_{-\bs{\kappa}_1} = 
V^{32}_{-\bs{\kappa}_1} = V^{13}_{-\bs{\kappa}_1}  = - V, \quad V^{\alpha\alpha}_{-\bs{\kappa}_2} =  -V e^{i\,\,2\pi\alpha/3}
\label{eq:phases}
\end{eqnarray}
then the lowest energy band will have integrated Berry curvature of $2\pi/3$ in each of the plaquettes in reciprocal space,
so Chern number ${\cal C} =1$. Although this result applies in the weak lattice limit $|V^{\alpha'\alpha}_{\kappa}| \ll E_{\rm r}$, the fact that the Chern number is a topological invariant guarantees robustness of the result up to moderate  $|V^{\alpha'\alpha}_{\kappa}|$.

Our considerations of the energy bands, and their topological properties, have been presented in reciprocal space. How do these considerations relate to the real-space magnetic field discussed in \S\ref{subsec:raman}? The magnetic field is defined in the limit in which the particle moves adiabatically in real space, valid when  $|V^{\alpha'\alpha}_{\bs{\kappa}}| \gg E_{\rm r}$. In this limit,  the kinetic energy term in (\ref{Goldman_eq23}) is negligible and the energy eigenstates follow from a tight-binding  model on the reciprocal space lattice. An elegant duality emerges \cite{coopermoessner}: the magnetic flux through the unit cell in real space due to adiabatic motion in the lowest energy dressed state is equal to the Chern number of the lowest energy band of the tight-binding  model defined by the couplings $V^{\alpha'\alpha}_{\bs{\kappa}}$ on the reciprocal space lattice. To form an ``optical flux lattice", with non-zero magnetic flux for the lowest energy dressed state through the real-space unit cell, one should just choose the couplings  $V^{\alpha'\alpha}_{\bs{\kappa}}$ on the reciprocal space lattice to give a lowest energy band with non-zero Chern number.
For the example of Fig.~\ref{Goldman_fig5}(a), with the above  couplings  (\ref{eq:phases}), the reciprocal space model is precisely the Harper-Hofstadter model for flux $1/3$ per plaquette. As discussed in Section \ref{section:models}, this model readily leads to Chern bands: for flux $1/3$ the lowest band has unit Chern number, so the couplings  (\ref{eq:phases}) form an optical flux lattice with one magnetic flux quantum through each real-space unit cell.

The non-zero flux densities of optical flux lattices lead to low-energy bands that are very similar to those
of the lowest Landau level: with unit Chern number, and with narrow width in energy. The bands depart from the exact degeneracy of Landau levels owing to the fact that the magnetic field and the scalar potential are non-uniform in space. However, for well-chosen parameters these effects can be made small in practical implementations \cite{cooperdalibard,cooperdalibard2013}.

\section{Discussion}\label{Dalibard_Discussion}

In this Chapter, we discussed how Chern bands can be created in cold-atom systems, through a wide variety of schemes based on atom-light coupling, rotation, or time-periodic modulations. Several of these techniques have been experimentally implemented in different laboratories, demonstrating  physical manifestations of synthetic magnetic fields  \cite{SchweikhardCEMC92,spielmanfield,Struck2011,LeBlanc2012,Struck:2013,Aidelsburger:2011,Aidelsburger:2013,Ketterle:2013,Aidelsburger:2014,Jotzu:2014}. 
Beyond the signatures discussed here, topological energy bands could also be probed in cold-atom systems  through the detection of Skyrmion patterns in time-of-flight images \cite{Alba:2011,Goldman:2013}, interferometry \cite{Abanin:2012,Duca:2015}, or the presence of chiral edge-states \cite{Stanescu2010,Goldman:2012prl,Buchhold:2012,GoldmanDalibard:2012}.

This Chapter focused on single-particle phenomena in Chern bands.  
Theoretical studies have extended the ideas presented here in a
variety of ways.  The methods for generating Chern bands
for cold atoms have been extended to higher dimensions and other symmetry classes,
including the $\mathbb{Z}_2$ topological insulator in three
dimensions \cite{hasankane,qizhang} both in
lattice-based  \cite{goldmantopoins,bermudez} and in
continuum \cite{bericooperz2} formulations, and to systems with
sublattice (chiral) symmetry\cite{essingurarie}. It is very
interesting to consider the effects of interactions on degenerate
bosons or fermions in the flat, or nearly flat, topological bands that we have described above.
For the Harper-Hofstadter model, numerical studies have established the existence of fractional
quantum Hall states, including the Laughlin
state \cite{sorensen:086803,palmer:180407,mollercooper-cf,PhysRevB.86.165314}
as well as states that exist only on the
lattice \cite{mollercooper-cf,PhysRevLett.108.256809}. The use of internal states to form
optical flux lattices leads to very flat topological bands which have
been shown to give rise to exotic non-Abelian phases of bosons even
for weak two-body repulsion \cite{cooperdalibard2013,PhysRevB.91.035115}, as well as to
interesting forms of ferromagnetic-nematic ordering for one-component
fermions \cite{baurcooper}. Studies of fractional Chern insulators have identified a
wealth of tight-binding lattice geometries with Chern bands where strongly correlated phases
can appear \cite{Parameswaran2013816,bergholtzliu} and which may find realizations in cold atom set-ups \cite{yaofci}. These theoretical works show that cold atomic gases have the potential to provide an ideal setting in which to explore novel forms of strongly correlated quantum phases, including fractional Chern insulators and other lattice-based strongly correlated topological phases.

\bibliography{Goldman_CUP_References}
\bibliographystyle{cambridgeauthordate}

\end{document}